\documentclass[twocolumn,showpacs12,preprintnumbers,amsmath,amssymb,superscriptaddress]{revtex4}

\usepackage{graphicx,epsfig}
\usepackage{bm}

\begin{document}

\title{\bf  A COMPARATIVE STUDY OF TWO MODELS OF QGP-FIREBALL FORMATION. }

\author{ R. Ramanathan$^1$, Y. K. Mathur$^1$, K. K. Gupta$^+$, \\  Agam K. Jha$^1$ and S. S. Singh}
\email{drramanathan@vsnl.net}

 \affiliation{ Department of Physics, University of Delhi, Delhi - 110007,INDIA  \\ $^+$Department of Physics, Ramjas College, University of Delhi, \\ Delhi - 110007, INDIA}

\begin{abstract}
\large  A Comparative study of the strengths and weakness of the models of fireball formation namely the statistical model of Ramanathan et.al (Physical Review C 70, 027903, 2004) and the approximation schemes of Kapusta et. al (Physical Review D 46, 1379, 1992) and its subsequent improved variants is made. The way to complement the various approximation schemes, in order to enhance their utility in the phenomenological analysis of QGP data that are expected from ongoing URHIC experiments, is suggested. The calculations, using the former model, demonstrate a striking QCD behaviour of the surface tension of the QGP droplet resulting in its increase with temperature, which is due to the confining nature of QCD forces at the surface and the interface surface tension varies as the cube of the transition temperature which is in conformity with the results of Lattice QCD simulations.\\

\end {abstract}

\maketitle

\large  The formation of QGP droplet (fireball) is one of the most exciting possibility in the ultra relativistic heavy ion collision (URHIC) [1]. The physics of such an event is very complicated and to extract meaningful results from a rigorous use of QCD appropriate to this physical system is almost intractable though heroic efforts at lattice estimation of the problem has been going on for quite some time [2]. One way out is to replicate the approximation schemes which have served as theoretical tools in understanding equally complicated atomic and nuclear systems in atomic and nuclear physics in the context of QGP droplet formation. This approach lays no claim to rigour or ab-initio ``understanding'' of the phenomenon but lays the framework on which more rigorous structures may be built depending on the phenomenological success of the model as and when testable data emerges from ongoing experiments.

In this paper we discuss the strength and weakness of two approximation schemes which seem promising in their usefulness in the above context and also explore their parametric interrelations which may be useful in phenomenological applications.The two approximation schemes under consideration are the QCD oriented QGP- droplet formation model of Csernai and Kapusta and its later refinements [3] and the simple statistical model of Ramanathan et. al [4] which again is essentially  QCD oriented in that use is made of an effective QCD potential. In both these schemes it is possible to estimate the fireball radius, the transition temperatures for their formation, nucleation rate etc. in quantitative terms for comparision with raw data as and when it is available from ongoing URHIC experiments.

As we shall see the two approaches can complement each other in the event of analysing the data on fireball formation especially with regard to nucleation of QGP droplets in a hadronic medium.

Central to this comparision is the rate $I$ to nucleate droplets of QGP in a hadronic gas per unit time per unit volume is given by [5] 

\begin{equation}
 I=I_{0}~e^{-\Delta F_{c}/T_{c}}
\end{equation} 

Where $I_{0}$ is the nucleation rate at vanishing change in free energy $ \Delta{ F} $ of the system due to the formation of a single critical size droplet of plasma. In the Kapusta et. al [3] formulation it is not possible to estimate the whole prefactor in terms of other possible measurable parameters of the droplet. 

The nucleation process is driven by statistical fluctuations being determined by the critical free energy difference between two phases. The Kapusta et. al model [3] uses the liquid drop model expansion for this , as given by

\begin{eqnarray} 
\Delta F = \frac {4\pi}{3} R^{3} [P_ {had}(T,\mu_{B}) - P_{q,g}(T, \mu_{B})] \nonumber \\
  + 4\pi R^{2} \sigma +\tau_{crit} T~ln \biggl [ 1 + (\frac {4 \pi}{3})R^{3}s_{q,g} \biggl]                                                                     \end{eqnarray}   

 The first term represents the volume contribution, the second term is the surface contribution where $\sigma$ is the surface tension, and the last term is the so called shape contribution . The shape contribution is an entropy term on account of fluctuations in droplet shape which we may ignore in the lowest order approximation. The critical radius $ R_{c}$ can be obtained by minimising (2) with respect to the droplet radius $R$ , which in the Linde approximation [6] is,

 \begin{equation}
 R_{c} = \frac {2\sigma}{\Delta p}~ or~                                             \sigma = \frac {3\Delta F_{c}}{4\pi R_{c}^{2}}
\end{equation}

There is no way to estimate the value of the crucial input $\sigma$ within the Kapusta et. al approach and the related approaches, though a value of $50~MeV/fm^{2}$ is assumed. The critical free energy in this model varies as in fig.$1$ [3].
\begin{figure}
\begin{center}
\epsfig{figure=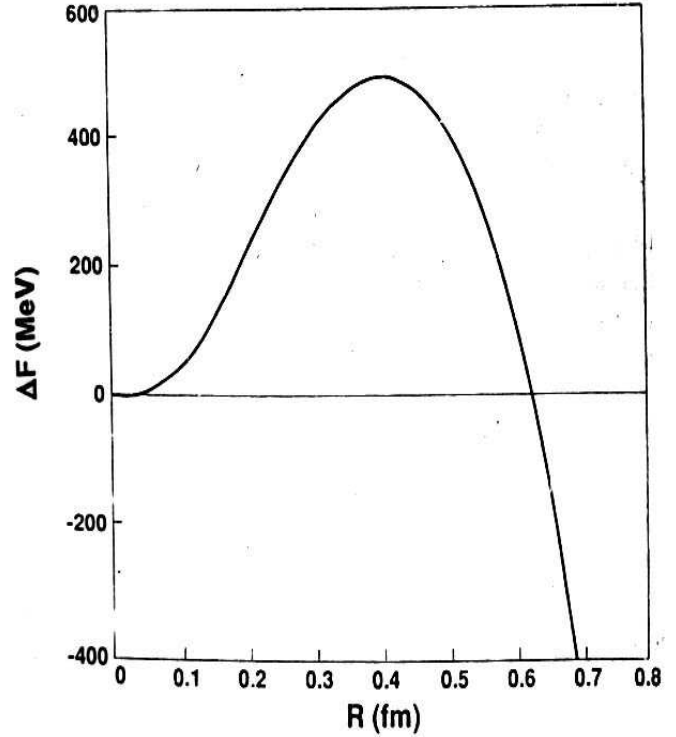,height=4.0in,width=3.5in}
\label{figure1}
\caption{\large The free energy difference $\Delta F(R)$ between a hadronic phase with and without a quark-gluon plasma droplet (PRC 51, 901; 1995)}
\end{center}
\end{figure}

It is easy to observe that this is a typical first order phase transition behaviour indicated by the Ramanathan et.al approach, thereby allowing direct comparision and the opportunity to use the calculational advantages of each of the approaches to bolster the usability of the approximation schemes as a phenomenological tool in the analysis of the data as we shall see in the following.

In the approximation schemes of the Ramanathan et. al [4], the relativistic density of states for the quarks and gluons is constructed adapting the procedures of the Thomas-Fermi construction of the electronic density of states for complex atoms and the Bethe density of states [6] for nucleons in the complex nuclei as templates. The expression for the density of states for the quarks and gluons (q, g)in this model is   

\begin{equation}
\rho_{q,g}(k) = (v / \pi^2)\biggl[(-V_{conf}(k))^2 \biggl(\frac{dV_{conf}(k)}{dk}\biggl)\biggl]_{q,g}
\end{equation}  

where $k$ is the relativistic four-momentum of the quarks and gluons, $v$ is the volume of the fireball taken to be a constant in the first approximation and$V_{conf}$ is a suitable confining potential relevant to the current quarks and gluons in the QGP [4] given by

\begin{equation}
[V_{\mbox{conf}}(k)]_{q,g} = (1/2k)\gamma_{q,g} ~ g^2 (k) T^2 - m_0^2 / 2k
\end{equation}

where $m_{0}$ is the mass of the quark which we take as zero for the up and down quarks and $150 MeV$ for the strange quarks. The $g(k)$ is the QCD running coupling constant given by 

\begin{equation}
g^2(k) = (4/3) (12\pi/27) (1/ ln(1+k^2/\Lambda^2))
\end{equation}                                                                       

where $\Lambda $ is the QCD scale taken to be $150~ MeV$.

The model has a natural low energy cut-off at 

\begin{equation}
  k_{min} = (\gamma_{q,g} N^{1/3} T^2 \Lambda^{2}/2)^{1/4}
\end{equation}                                                                                                                               
with $$N = (4/3)(12\pi/27).$$
The freee energy of the respective case $i$ (quarks, gluons, interface etc.) for Fermions and Bosons (upper sign or lower sign)can be computed using the following expression

 \begin{equation}
F_i = \mp T g_i \int dk \rho_i (k) \ln (1 \pm e^{-(\sqrt{m_{i}^2 + k^2}) /T})
\end{equation}                                                                       
with the surface free-energy given by a modified Weyl [7] expression:

\begin{equation}
F_{surface} = \frac{1}{4}  R^{2}  T^{3} \gamma
\end{equation}                                                                       
where the hydrodynamical flow parameter for the surface is :

\begin{equation}
\gamma = \sqrt{2}\times \sqrt{(1/\gamma_g)^2+(1 / \gamma_q)^2},
\end{equation}                                                                    
For the pion which for simplicity represents the hadronic medium in which the fireball resides, the free energy is :

 \begin{equation}
F_{\pi} = (3 ~ T/2\pi^2 )(-v) \int_0^{\infty} k^2 dk \ln (1 - e^{-\sqrt{m_{\pi}^2 + k^2} / T})
\end{equation}

With these ingredients we can compute the free-energy change with respect to both the droplet radius and temperature to get a physical picture of the fireball formation, the nucleation rate governing the droplet formation, the nature of the phase transition etc. This can be done over a whole range of flow-parameter values [4], We exhibit only the two most promising scenarios in fig.$2$ and fig.$3$.

\begin{figure}
\begin{center}
\epsfig{figure=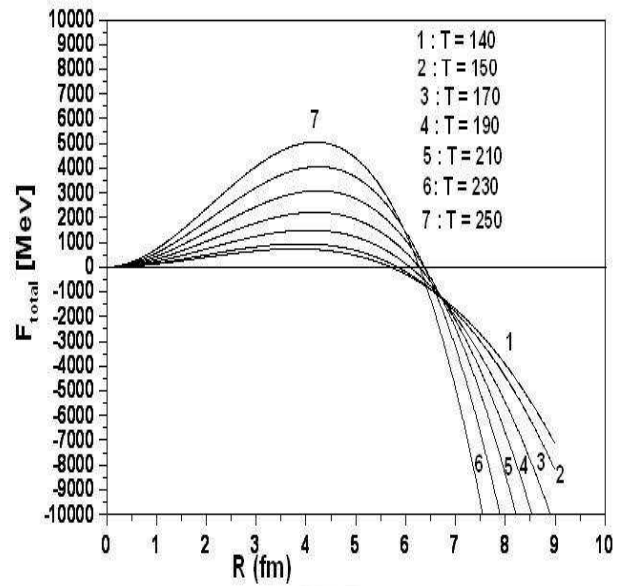,height=4.0in,width=3.5in}
\label{fig4}
\caption{\large $F_{total}$ at $\gamma_{g} = 6\gamma_{q}$, $ \gamma_{q} = 1/6 $ for various temperatures.}
\end{center}
\end{figure}

\begin{figure}
\begin{center}
\epsfig{figure=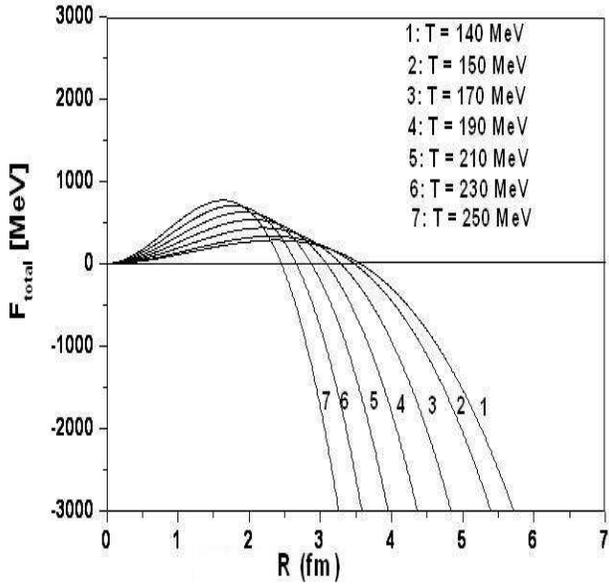,height=4.0in,width=3.5in}
\label{fig5}
\caption{\large $F_{total}$ at $\gamma_{g} = 8\gamma_{q}$ , $ \gamma_{q} = 1/6 $ for various temperatures.}
\end{center}
\end{figure}

As could be seen the general nature of the curves given by fig.$1$ and fig.$2$ and $3$, though using two totally different approches leads to the conclusion that both predict a first order phase transition for the fireball production process. As a derivative computation we can also compute the nucleation rate leading to the droplet formation, which for the two scenarios is illustrated by figs.$ 4$  and $5$.

\begin{figure}
\begin{center}
\epsfig{figure=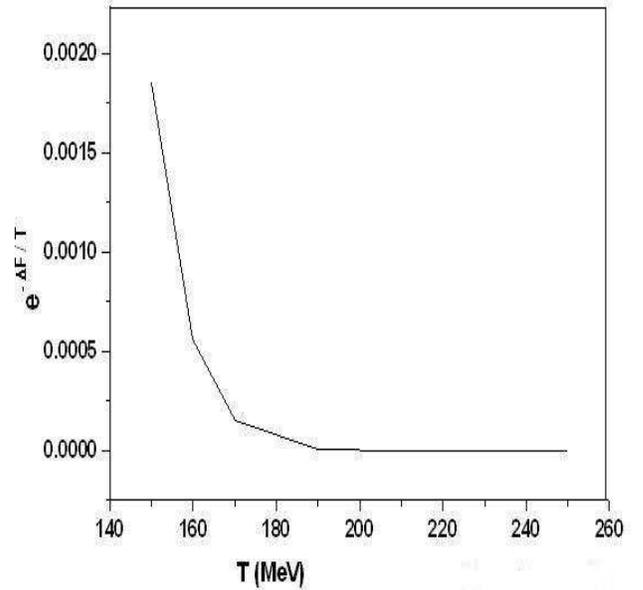,height=4.0in,width=3.5in}
\label{exp2}
\caption{\large The fireball formation rate with temperatures  at $\gamma_{g} = 6\gamma_{q}$, $ \gamma_{q} = 1/6 $ .}
\end{center}
\end{figure}

\begin{figure}
\begin{center}
\epsfig{figure=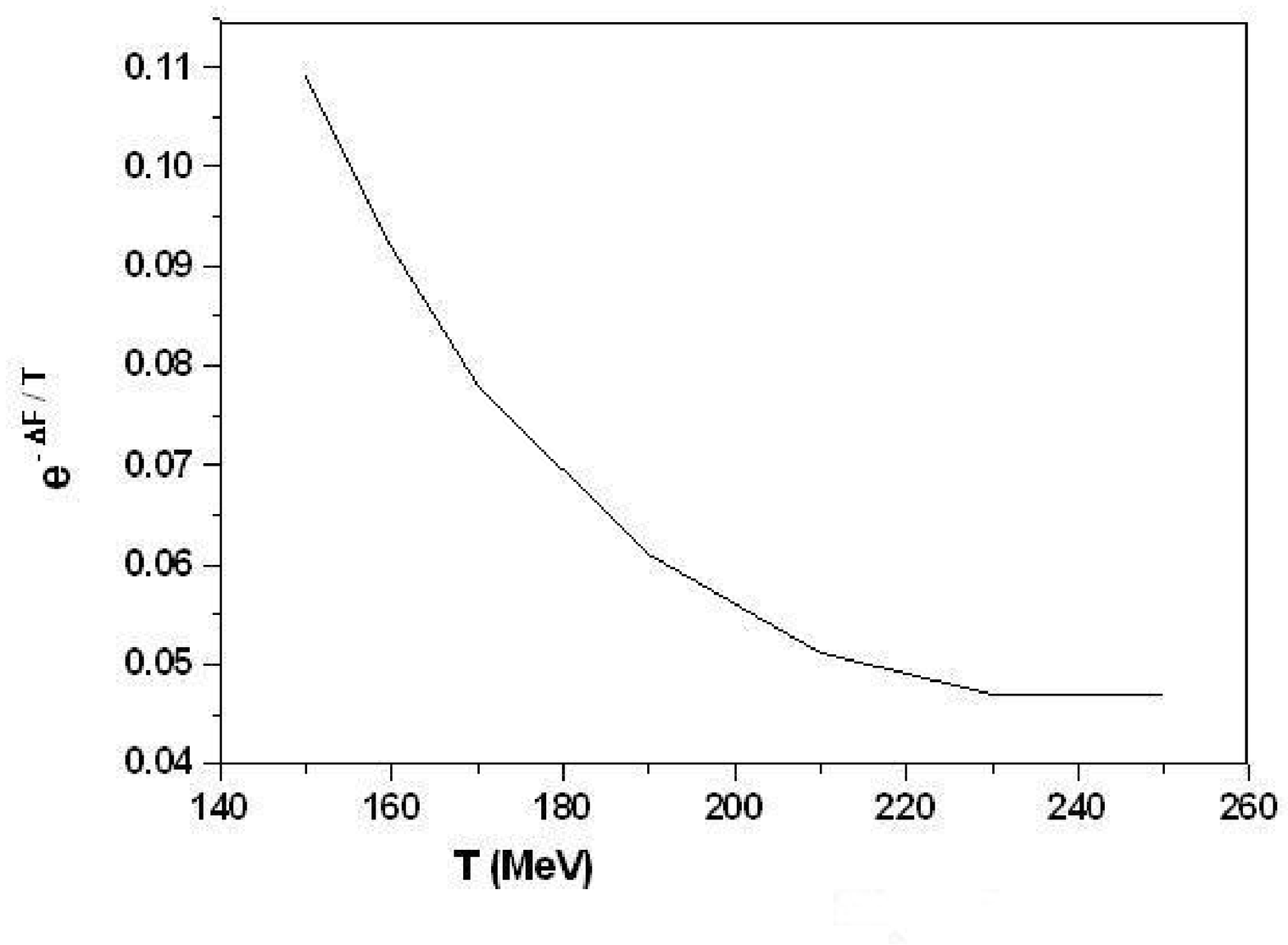,height=4.0in,width=3.5in}
\label{exp3}
\caption{\large The fireball formation rate with temperatures at $\gamma_{g} = 8\gamma_{q}$ , $ \gamma_{q} = 1/6 $ .}
\end{center}
\end{figure}

As could easily be observed from figs 4 and 5, for both the sets of parameter values, the droplet formation rate grows exponentially in the vicinity of transition temperature range of $150$ to $170~ Mev$ as expected from lattice simulations and other model calculations. The critical free-energy and radius at the maximum of the curves in figs. 2 and 3 allow us to compute the surface tension used in the Kapusta et.al and the derivative models [3].
\vfill
\begin{quote}
\begin{tabular}{|r|r|r|r|r|}
\hline
$T_{c}$&$\Delta F_{c}$&$R_{c}$&$\sigma$&$\frac{\sigma}{T_{c}^{3}}$\\  
$(MeV)$&$(MeV)$&$(fm)$&$(MeV/fm^{2})$&\\
\hline
150&332.203&3.475&6.568&0.078\\
160&382.359&3.385&7.966&0.078\\
170&433.037&3.285&9.580&0.078\\
190&532.219&3.085&13.35&0.078\\
210&623.349&2.875&18.004&0.078\\
230&702.041&2.655&23.776&0.078\\
250&766.041&2.455&30.343&0.078\\
\hline
\end{tabular}
\end{quote}
Table-1 for Surface Tension of QGP droplet at $\gamma_{g}=8\gamma_{q}$,$ \gamma_{q} = 1/6$.
\vfill

\begin{quote}
\begin{tabular}{|r|r|r|r|r|}
\hline
$T_{c}$&$\Delta F_{c}$&$R_{c}$&$\sigma$&$\frac{\sigma}{T_{c}^{3}}$\\
$(MeV)$&$(MeV)$&$(fm)$&$(MeV/fm^{2})$&\\
\hline
150&943.595&5.835&6.616&0.078\\
160&1197&5.965&8.031&0.078\\
170&1494&6.085&9.633&0.078\\
190&2216&6.275&13.435&0.078\\                                               
210&3088&6.375&18.140&0.078\\                                                 
230&4059&6.375&23.844&0.078\\
250&5052&6.275&30.630&0.078\\
\hline
\end{tabular}                                       
\end{quote}
Table-2 for Surface Tension of QGP droplet at $\gamma_{g}=6\gamma_{q}$,$ \gamma_{q} = 1/6$.                                                                     \vfill

 The tables $1$ and $2$ list the surface tension computed in the Ramanathan et.al model which can be used in the dynamical models[3], thus reducing the arbitrariness in the latter models to this extent, thus enabling us to use the parameter extracted from a static situation to make perdictions about the dynamical growth process of fireball formation.      

In both tables $1$ and $2$ the surface tension is seen to increase with the temperature of the fireball,which is a beautiful demonstration of a QCD effect.As the temperature of the QGP droplet increases the shear forces on the fireball surface will also increase tending to tear the surface quarks apart, consequently, bringing into play the confining property of the QCD forces manifesting itself in increased surface tension, which is exactly what the calculations show.Another striking feature of the result is the independence of the QGP droplet surface tension $\sigma$ variations in the flow parameters of the model and it varies with only the temperature, in the lowest order approximation we have employed.

The constancy of the ratio $\frac {\sigma}{T_{c}^{3}}$ indicates a cubic crtitical temperature dependence of the surface tension of the interfacial separation between the two phases. This is in striking conformity with the results of Lattice QCD simulation [8]. 

{\bf Acknowledgement:}
  
  We are very thankful to Prof. R. Venugopalan for constructive suggestions and discussions.     

{\bf References :}
\begin{enumerate}
\item{F. Karsch, E. Laermann, A. Peikert, Ch. Schmidt and S. Stickan, Nucl. Phys. B(Proc. Suppl.)94, 411 (2001)}.
\item{T. Renk, R. Schneider, and W. Weise, Phys. Rev. C66, 014902 (2002)}.
\item{L. P. Csernai, J. I. Kapusta, E. Osnes, Phys. Rev. D67, 045003 (2003), e-Print Archive: hep-th/0201024; J. I. Kapusta, R. Venugopalan, A. P. Vischer, Phys. Rev. C51, 901-910 (1995), e-Print Archive: nucl-th/9408029; L. P. Csernai, J. I. Kapusta, Phys. Rev. D46, 1379-1390 (1992). P. Shukla and A. K. Mohanthy, Phys. Rev. C 64, 054910(2001)}.
\item{R. Ramanathan, Y. K. Mathur, K. K. Gupta, and Agam K. Jha, Phys. Rev. C 70, 027903 (2004); R. Ramanathan, Y. K. Mathur, K. K. Gupta, Proc. IVth Int. Conf. on QGP, Jaipur (2001)(unpublished);R. Ramanathan, Y. K. Mathur, K. K. Gupta and Agam K. Jha, e-Print Archive:  hep-ph-0402272 (2004)}.
\item{  B. D. Malhotra and R. Ramanathan, Phys. Lett. A 108, 153,1985; See in references (3)}.
  \item{E. Fermi, Zeit F. Physik 48, 73 (1928); L. H. Thomas , Proc. camb. Phil. Soc. 23, 542 (1927);H. A. Bethe, Rev. Mod. Phys. 9, 69 (1937)}.
\item{H. Weyl, Nachr. Akad. Wiss Gottingen 110 (1911)}.
\item{Y. Iwasaki, K. Kanaya, Leo Karkkainen, K. Rummukainen, and T. Yoshie, e-print arXiv: hep-lat/9309003; M. Hasenbusch, K. Rummukainen and K. Pinn, e-print arXiv: hep-lat/9312078}. 
\end{enumerate}

\end{document}